\documentstyle[prl,aps,multicol,psfig,rotate,citesort]{revtex}

\begin{document}

\title{Comparative simulation study of colloidal gels and glasses}
\author{Antonio M. Puertas~\cite{puertas}, Matthias Fuchs
and Michael E. Cates}

\address{Department of Physics and Astronomy, The University of
Edinburgh, EH9 3JZ, UK}
\date{\today}
\maketitle

\begin{abstract}
Using computer simulations, we identify the mechanisms causing
aggregation and structural arrest of colloidal suspensions interacting with
a short-ranged attraction at moderate and high densities.
Two different non-ergodicity
transitions are observed.  As the density is increased, a
glass transition takes place, driven by excluded volume effects. In
contrast, at moderate densities, gelation is
approached as the strength of the attraction increases. 
At high density and interaction strength, both transitions merge, and a
logarithmic decay in the correlation function is observed. All of these
features are correctly predicted by mode coupling theory.
\end{abstract}


\begin{multicols}{2}

Colloidal dispersions aggregate into various non-equilibrium structures
depending on density, interaction strength and
range. The accompanying rheology and structure are among the
key properties desired for their technological applications
\cite{Russel89}. Moreover, thanks to the possibility to tailor
effective  interactions by e.g. addition of salt and  polymer,
colloids allow us to study the fundamental mechanisms of kinetic
arrest. Whereas colloidal hard spheres have become a
model system for the study of structural arrest at a glass transition
\cite{Megen98}, colloidal gelation has only recently been
associated with glassy behavior \cite{Verduin95,Bergenholtz99,Segre01}.
Colloidal gelation is ubiquitous in suspensions driven by attractions 
of quite short range
and moderate to high strength \cite{Poon97}. At low packing fractions, it
entails the formation of heterogeneous and often
self-similar networks; there, an interplay of phase separation kinetics and
percolation often are considered responsible for its existence
\cite{Poon97}.  At higher densities, the gelation boundary
extends into the homogeneous fluid region \cite{Grant93,Verduin95},
where it also lies well separated from estimates of percolation 
\cite{Grant93,Verduin95,Rueb98,Mallamace00}.  Crossing into the gelled
state anywhere along the transition line results in qualitatively
the same phenomena, like flow properties that indicate solidification
\cite{Grant93,Rueb98}, and non-ergodic dynamics according to light
scattering  \cite{Verduin95,Poon99,Pham01}. 

We present simulations designed to identify the mechanism of
colloidal gelation driven by attractions of only moderate
strength. Because of the distance of the gel boundary from other  
boundaries (percolation and phase separation) at higher
densities, we concentrate on these, where we sweep out the
region between gel and glass transition lines. We show that both 
non-equilibrium transitions are caused by a slowing down of local
rearrangements, as predicted well by mode coupling theory
(MCT)  \cite{Fabbian99,Bergenholtz99,Dawson01}. We contrast the glass
transition, caused by caging of particles owing to steric hindrance,
with attraction-driven gelation caused by bonding
between particles. We verify that the simultaneous presence
of two non-ergodic states results in anomalous non-exponential
(logarithmic) time dependences, as recently conjectured to
explain observations in micellar systems
\cite{Mallamace00} or microgel suspensions \cite{Bartsch94}.

The simulated system comprises 1000 soft-core ($V(r=|{\bf r}_{i}-{\bf r}_j|)
\propto(a_{ij}/r)^{36}$, $a_{ij}=a_i+a_j$) particles of mean 
radius $a$ with
polydispersity in size (flat distribution with 10\% width) to prevent
crystallisation. Densities are reported as packing fractions
$\phi_c=\frac{4\pi}{3} n a^3$.
A short range attraction, mimicking the polymer
induced depletion attraction in experimental systems 
\cite{Russel89,Poon97,Poon99,Pham01}, is given by an
Asakura-Oosawa (AO) form generalized to polydisperse systems
\cite{mendez00,potshape}. The range of the attraction, $2 \xi$, is 
set to $0.2 a$, and its strength is
proportional to the polymer concentration $\phi_p$.
To help avoid liquid-gas separation, a weak long range barrier is added to
the potential. The barrier extends from $a_{12}+2\xi$ to $4a$, and is
described by a fourth order polynomial matched to give a continuous force.
Its maximal height is $1k_BT$, which equals the 
depth at contact of the AO potential at $\phi_p=0.0625$. 
In all states studied, the barrier is much smaller than the attraction, and
in the purely repulsive case ($\phi_p=0$) it is omitted.
We will measure lengths in units of $a$ and
time in units of $\sqrt{4a^2/3v^2}$, where the thermal velocity,
$v$, is set to $2/\sqrt{3}$. Equations of motion were integrated using the 
velocity-Verlet algorithm, with a time step of $0.0025$. Colloidal 
dynamics (neglecting hydrodynamic interactions) were mimicked by running 
the simulations in the canonical 
(constant NTV) ensemble, where the thermostat plays the role 
of the surrounding liquid. Every $N$ time steps, the velocity of the particles 
was rescaled to assure constant $v$. No effect of $N$ on the results 
was observed for well equilibrated samples.

The central quantity of our study will be  the self part of the
intermediate scattering 
function, $\Phi_q^s(t)=\langle\exp i {\bf q}\cdot \left( {\bf r}_j(t)-
{\bf r}_j(0) \right) \rangle$, for wave-vector ${\bf q}$, where $\langle 
\dots \rangle$
denotes an average over particles and time origin. 
$\Phi_q^s(t)$ allows us to probe and identify the nature of the dominant
dynamical mechanism because of $(i)$ its $q$-dependence and $(ii)$
the detailed predictions that are available from MCT.
Indeed, if a structural arrest at a non-ergodicity transition is
approached, $\Phi_q^s(t)$ reveals a two-step process, where 
the decay  from the plateau is given by the von Schweidler power-law
series \cite{Franosch97}:
\begin{equation}
\Phi_q^s(t)\:=\:f_q^s\,-\,h_q^{(1)} (t/\tau)^b\,+\,h_q^{(2)}
(t/\tau)^{2b}\,+\,O(t^{3b}) \; .
\label{alpha-decay} \end{equation}
Here $f_q^s$ is the non-ergodicity parameter, $h_q^{(1)}$ and $h_q^{(2)}$ 
are amplitudes, and $b$ is known as the von Schweidler
exponent. On the one hand, the observation of this (universal) von Schweidler
behavior -- and tests of further relations, as done below --
establishes that a feed-back mechanism in the structural relaxation causes
arrest. On the other hand, the (non-universal)
wave-vector dependence of the amplitudes, like $f_q^s$, allows us to
identify the specific kinetic process which freezes out. As the 
transition is approached, the characteristic time $\tau$ diverges as $\tau
\propto |\phi-\phi^c|^{-\gamma}$, where $\gamma$ is determined by $b$, see
e.g. \cite{Franosch97}. 
\begin{figure}[htb]
\psfig{file=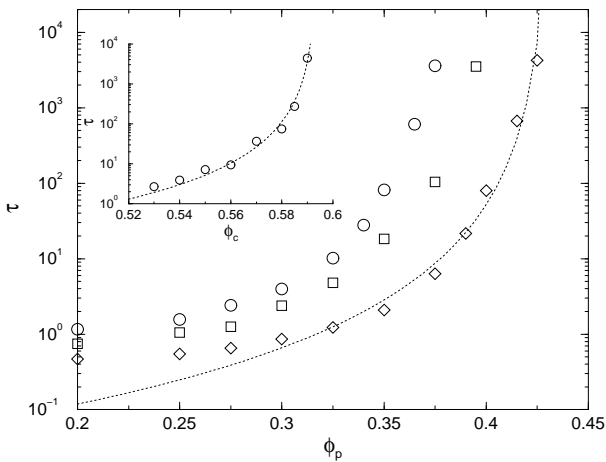,width=3.2in}
\caption {Relaxation time $\tau$ as a function of $\phi_p$ for three 
colloid volume fractions: $\phi_c=0.40\,(\Diamond)$, $\phi_c=0.50\,(\Box)$,
$\phi_c=0.55\,(\bigcirc)$. Inset: $\tau$ vs. $\phi_c$ for soft spheres
($\phi_p=0$, $V(r)\sim r^{-36}$). Dotted lines are fittings with 
predetermined $\gamma$.}
\label{fig1} 
\end{figure}

Figure 1 presents evidence for both the repulsion and attraction driven glass
transitions, as identified by a diverging $\tau$ \cite{taucom}. Upon increasing 
the packing fraction $\phi_c$ (inset in figure 1), the system 
approaches a glass transition caused by steric hindrance, which we have
studied including only the $r^{-36}$ repulsion ($\phi_p=0$ and no barrier)
\cite{barrier}. The transition correlates well
with observations at the colloidal glass transition \cite{Megen98}
and previous simulations of e.~g. a glassy Lennard-Jones mixture
\cite{Kob}. We have analysed it using the concepts of idealized MCT, but
will present only a few results here for comparison with gelation.
Gelation itself is induced by strengthening the attraction at
intermediate packing fractions (figure 1). There,  
far from equilibrium or percolation transitions
(as we tested by monitoring the static structure factor),
arrest again is of kinetic origin, and occurs at lower 
attraction strengths the higher $\phi_c$.
\begin{figure}[htb]
\psfig{file=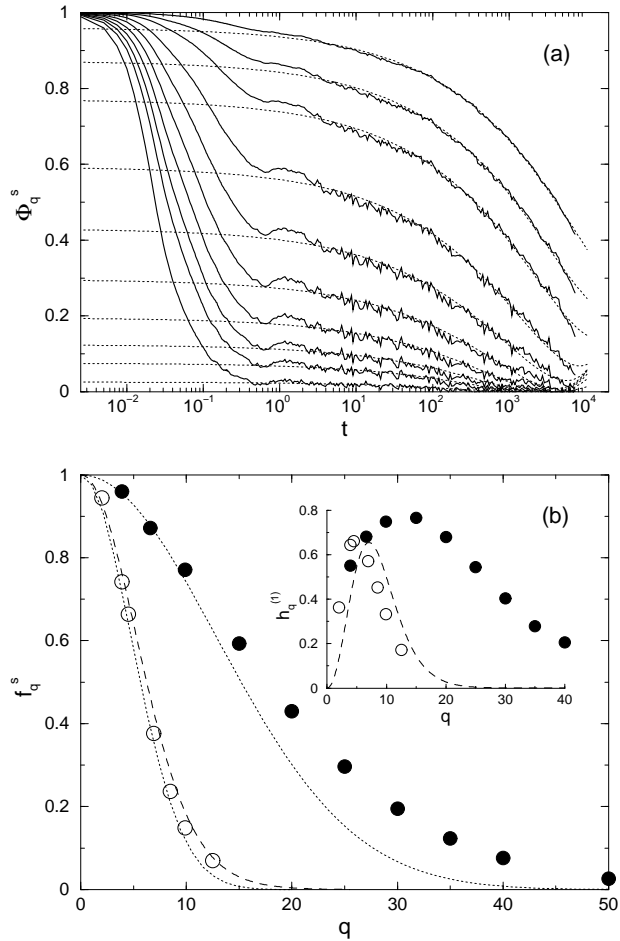,width=3.2in}
\caption { (a): Correlation functions at 
$\phi_c=0.4$, $\phi_p=0.425$ and von Schweidler fits. From top to botton, 
$q=3.9,\,6.9,\,9.9,\,15,\,20,\,25,\,30,\,35,\,40,\,50$. (b): $f_q^s$ for 
glass ($\circ$) and gel ($\bullet$) (from (a)) transitions, 
with the Gaussian
approximation for both of them (dotted lines) and the MCT result
for hard spheres (dashed line) \protect\cite{Fuchs92}.
Inset: $h_q^{(1)}$ using $\tau$ from figure 1, and MCT result
for hard spheres.} \label{fig2}
\end{figure}

To shed light on this transition, the
correlation functions at different wave-vectors were studied.
 The slowest state at $\phi_c=0.40$ is presented in Fig \ref{fig2}. 
The upper panel shows the self intermediate scattering functions for
different wave-vectors, and the fits using 
(\ref{alpha-decay}) up to second order. A common exponent $b$ was taken in
the fitting, yielding a value of $b=0.38$, appreciably 
lower than the hard spheres value $b=0.53$ (which we found for our
soft sphere glass at $\phi_p=0$).  As predicted by MCT, we
can calculate from $b$ the divergence of the relaxation times in
Fig. \ref{fig1}. The resulting value $\gamma=3.03$ fits  
the data, while $\gamma=2.63$ at the soft sphere glass
transition \cite{barrier}. The gel transition is estimated to occur at
$\phi_p=0.431$.  Because we find the universal properties predicted by
MCT, we conclude that at $\phi_c=0.40$ the gel transition is a
regular non-ergodicity transition in the structural dynamics. 

Their very different $q$-width (Fig. \ref{fig2}b) for the
non-ergodicity parameters and amplitudes brings out a major 
difference in the two underlying mechanisms. 
Whereas repulsions localize the particle within a cage, which it can
explore up to mean squared displacements $r_l^2$ of the order of $r_l^2=0.13$ 
(from our simulations, not shown; $r_l^2=0.134$ from MCT 
\cite{Fuchs92}), attractions bind the
particle to its neighbors and thus localize it much more
tightly.  At $\phi_c=0.40$, we find $r_l^2=0.018$ by simulations, which is 
of the order of a low-density estimate
\cite{Bergenholtz99} for our interaction range, $2\xi=0.2$.
The corresponding high amplitudes $f_q^s$ of density fluctuations are 
consistent with light scattering observations at fixed $q$
\cite{Poon99,Pham01} and with MCT calculations
\cite{Bergenholtz99,Fabbian99,Dawson01}. These fluctuations extend to large
$q$ and relax only when the particles break free from their bonds.
The comparison with the Gaussian
approximation, $f^{s {\rm G}}_q=\exp{\{-q^2 r_l^2/6\}}$
evidences stronger non-Gaussian effects at gelation than at the glass
transition. We stress the cooperativity of the
structural relaxation at both transitions. Holding all particles fixed,
except for one, leads to mean squared displacement for the tracer (as it 
explores the frozen enviroment) much smaller than in the free system (before
the start of the structure relaxation of the free system, the ratio is
$\approx 6$ for both cases).
The cage or network of bonds around an arrested particle thus necessarily 
fluctuates with it.
\begin{figure}[htb]
\psfig{file=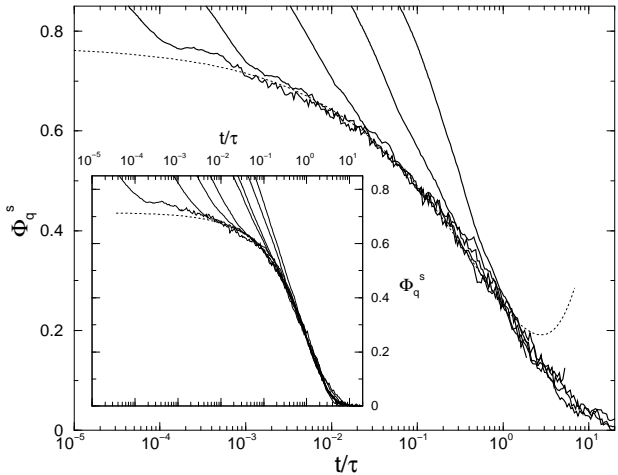,width=3.2in}
\caption {Correlation functions vs. rescaled time $t/\tau$ for
$\phi_c=0.4$ and $\phi_p=0.375,\,$$0.39,\,$$0.40,\,$$0.415,\,$$0.425$ 
(from right to left) and von Schweidler fit. Inset: Same plot for 
soft spheres and $\phi_c=0.53,\,$$0.54,\,$$0.55,\,$$0.56,\,$$0.57,\,$
$0.58,\,$$0.585,\,$$0.59$, and MCT master curve \protect\cite{Fuchs92}.}
\label{fig3}
\end{figure}

To test further the nature of the gel transition, the scaling of the
final (or $\alpha$-) decay was studied. In Fig. \ref{fig3} we
present the rescaled ($\Phi_q^s(t/\tau=1)=0.25$) correlation 
functions at $q=9.9$ for different attraction strengths, close to the gel 
transition. 
In the inset to this figure, a similar plot
deals with the glass transition ($q=3.9$ in this case). In both
cases, the curves clearly collapse during the $\alpha$-decay
indicating an unique mechanism which dominates the slowing
down at the transitions. For the purely repulsive case,  
the MCT master curve for the rescaled decay of hard spheres \cite{Fuchs92} at a
slightly larger  wave-vector ($q=4.3$) is also presented, confirming
the quantitative agreement between MCT and our results.
In the gel case, no master function is available, but the
fit to (\ref{alpha-decay}) is presented. The different
stretching in the two cases is clear.
\begin{figure}[htb]
\psfig{file=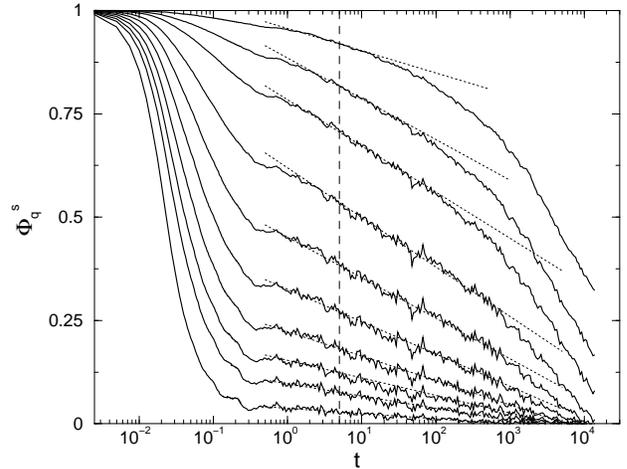,width=3.2in}
\caption{Same as figure 2(a) for $\phi_c=0.55$ and $\phi_p=0.375$.
Dotted lines: logarithmic fittings to the correlators around $f_q$. The
dashed line represents $t_1$. }\label{fig4}
\end{figure}

\begin{figure}[htb]
\psfig{file=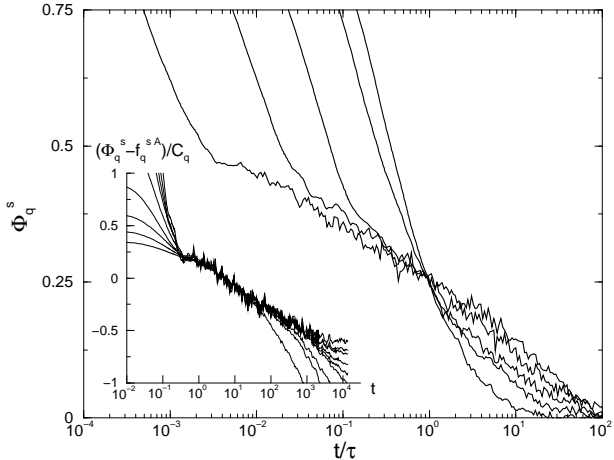,width=3.2in}
\caption {Correlation function vs. rescaled time, $t/\tau$, at $q=20$. 
$\phi_c=0.55$, and $\phi_p=0.325,\,0.34,\,0.35,\,0.365,\,0.375$ from 
right to left at $t<\tau$. Inset:
$(\Phi_q^s-f_q)/C_q$ as a function of time for the same state and
wave-vectors as figure \ref{fig4} (wave-vectors increasing from top to
botton).}\label{fig5}
\end{figure}

We study now the gel transition at a higher density $\phi_c=0.55$, where 
it lies closer to the glass transition. Within MCT, the simultaneous
existence of  two different non-ergodicity transition branches 
opens the possibility for
end-point singularities, where the branches merge \cite{gotze89}.
In systems with short range
attractions, such a singularity has been predicted 
close to the crossing of the two transition lines (the actual
distance depending on the details of the potential)
\cite{Fabbian99,Bergenholtz99,Dawson01}. Close to the singularity,  
a logarithmic decay around the plateau in the correlation function,
$\Phi_q^s(t)=f_q^{sA}-C_q \ln (t/t_1)$, is a proposed signature 
\cite{Fabbian99,Dawson01}, and intriguingly is observed experimentally
in more complicated systems \cite{Bartsch94,Mallamace00}. 
Having identified two different non-ergodicity transitions in
our system, we now test this prediction. In Fig. \ref{fig4},
 the correlation functions at the same wave-vectors
as in Fig. \ref{fig2} are presented for the state $\phi_p=0.375$. Logarithmic
decays are observed in all of the correlators (linear traces in the
plot) for up to three decades in time,
signalling a higher order singularity  nearby. 
It is interesting to note that the logarithmic trace of the correlator has
different extents, depending on the wave-vector. To make it clearer,
$(\Phi_q^s-f_q^{sA})/C_q$ vs. time has been plotted as an inset to
Fig. \ref{fig5}, where $f_q^{sA}$ is determined at $t_1=5$ (vertical 
line in Fig. \ref{fig4}). 
Deviations from the logarithmic decay are stronger, the higher
$f_q^s$. This is in complete agreement with the
theoretical expectations in \cite{Dawson01}. Since both the 
von Schweidler decay (associated with the gel transition) and the logarithmic 
trace now take place in
the same window, important corrections to the $\alpha$-scaling of the curves
are expected. This is seen in the main graph of
Fig. \ref{fig5}: the long time dynamics at
different states cannot  be collapsed onto a master curve by time rescaling.
This shows that two mechanisms are responsible for the structural
slowing down and that changes in the control parameters change the
relative distance to gelation but also to the higher order singularity.
The intermediate isochore, 
$\phi_c=0.50$, shows a mixed behavior: a logarithmic decay 
in a smaller window than for $\phi_c=0.55$, and followed by an apparent
 power law decay.

In summary, using MD simulations, we have deduced from the wave-vector
dependence of the dynamical density fluctuations, that repulsion and 
short-ranged attraction lead  to two different structural arrests at 
high enough density or attraction strength, respectively. At the merging  
of both glassy states, subtle logarithmic time
variations appear. Comparing with the recent MCT predictions of these 
phenomena we find perfect agreement.

We thank W. Kob for valuable discussions.
M.F.\ was supported by the Deutsche Forschungsgemeinschaft, grant Fu~309/3,
and A.M.P. by the Ministerio de Educaci\'on y Cultura.

\end{multicols}


\begin{thebibliography}{99}
\bibitem{potshape}
The potential minimum was set to $a_{12}$ using a parabola for 
$r<a_{12}+\xi/5$, smoothly connected to the AO.
\bibitem{taucom}
The relaxation time was defined by $\Phi_q^s(t=\tau)=0.25$ at
$q=3.9$ for the glass transition and $q=9.9$ for gelation.
\bibitem{barrier}
The glass transition was shifted from $\phi_c=0.594$ to $\phi_c=0.55$, upon
inclusion of the barrier. No qualitative differences were observed.
\end{thebibliography}
\end{document}